# Dual mechanisms for transient capacitance anomaly in improper ferroelectrics

*Xin Li,[1‡] Yu Yun,[1‡]*, Pratyush Buragohain[1], Haidong Lu[1], Arashdeep Singh Thind[2,3], Donald A. Walko[4], Detian Yang,[1] Rohan Mishra,[2,3] Alexei Gruverman[1,5], Xiaoshan Xu[1,5]**

[1]Department of Physics and Astronomy, University of Nebraska, Lincoln, Nebraska 68588, USA
[2]Institute of Materials Science & Engineering, Washington University in St. Louis, St. Louis MO, USA
[3]Department of Mechanical Engineering & Materials Science, Washington University in St. Louis, St. Louis MO, USA
[4]Advanced Photon Source, Argonne National Laboratory, Argonne, Illinois 60439, USA
[5]Nebraska Center for Materials and Nanoscience, University of Nebraska, Lincoln, Nebraska 68588, USA

‡These authors contributed equally to this work.
*Corresponding author: Xiaoshan Xu (X.X.), Yu Yun (Y.Y.)

**Abstract:**

Negative capacitance (NC) effects in ferroelectrics can potentially break fundamental limits of power dissipation known as 'Boltzmann tyranny'. However, the origin of transient NC of ferroelectrics, which is attributed to two different mechanisms involving free-energy landscape and nucleation, is under intense debate. Here, we report coexistence of transient NC and an "S"-shape anomaly during the switching of ferroelectric hexagonal ferrites capacitor in an RC circuit. The early-stage NC arises from the nucleation process, while the late-stage "S"-shape anomaly corresponds to a nascent NC associated with the free-energy landscape. The entire waveform can be reproduced using a hybrid model that simultaneously incorporates these two mechanisms. These results highlight the multi-variable free-energy landscape of hexagonal ferrites that enables abrupt change of internal field and demonstrate that the two mechanisms are not mutually exclusive, resolving the long-standing debate. The behavior of the "S" shape anomaly also provides a pathway to extract parameters of free-energy landscape and switching dynamics.

Ferroelectricity, originating from broken inversion symmetry of crystal structures, can be described using the polarization-dependent free-energy landscape in the phenomenological Landau theory. Spontaneous polarization states locate at energy minima (**Fig. 1(a)**) and the transition from ferroelectric to paraelectric phases corresponds to disappearance of these minima[1-4]. Intriguingly, the barriers between the energy minima can be exploited to boost the capacitance of ferroelectric/dielectric heterostructures, which is promising for enhancing the energy efficiency beyond the lower limit of power consumption in field-effect transistors[5].

In principle, free-energy landscape of ferroelectrics can be revealed by polarization switching processes[6-11], provided that polarization can be varied continuously. In this case, according to the Landau-Khalatnikov (LK) model[10,11], the energy barriers lead to a non-monotonic polarization switching speed and NC response (Section S1 within the Supplemental Material). Naturally, the recent observation of transient NC[12,13] has been attributed to the mechanism of free-energy landscape. However, LK model is unable to reproduce the transient NC quantitatively, even with multi-grain model[14] or time-dependent viscosity coefficient[15]. Moreover, polarization switching is known to normally involve complex processes like nucleation of reversed domains and domain-wall propagation[16-18]. Overcoming the nucleation energy barrier can also result in a non-monotonic polarization switching speed (**Fig. 1(b)**) and NC response (Section S2 within the Supplemental Material). Consequently, whether the transient NC originates from the mechanism of free-energy landscape[12,13,15,19] or the mechanism of nucleation remain under debate.

To resolve this debate, the ferroelectrics with both abrupt change of internal field from LK model ($E_{LK}$ in Fig.1(a)) and nucleation-involved polarization switching are necessary, so that resultant responses of transient capacitance with different origins can compare directly. Here, we focus on improper ferroelectrics hexagonal ferrites (h-$R$FeO$_3$, $R$: rare earth), the crystal structure of h-$R$FeO$_3$ consist of triangular lattice of FeO$_5$ bipyramids sandwiched by rare earth layers (**Fig. 1(c)**). Spontaneous polarization is induced by non-polar K$_3$ structural distortion[20-25], characterized by the in-plane displacement of apical oxygen described by the magnitude ($Q$) and angle (or phase) ($\phi$) (**Fig. 1(c)**). With three variables, $Q$, $\phi$, and polarization ($P$), a complex free-energy landscape is expected[26]. A two-dimensional ($Q$ and $\phi$) representation of this landscape, after minimization with respect to $P$, is displayed in **Fig. 1(d)**, where three minima at $\phi = 2n\frac{\pi}{3}$ and three at $\phi = (2n+1)\frac{\pi}{3}$ correspond to $P > 0$ and $P < 0$, respectively, $n$ is an integer [27]. Notably, the system maintains a constant $\phi$ during the polarization switching [see **Fig. 1(e)**] before the local energy minimum disappears, at which point a rapid change of $E_{LK}$ is expected, potentially leading to a more obvious NC from the mechanism of free-energy landscape [28].

In this work, we demonstrate that two NC mechanisms coexist and manifest in different timescale during the switching process of a hexagonal ferrite ferroelectric capacitor in a RC circuit. The mechanism of nucleation leads to a transient NC with the time scale of nucleation (~ 1 μs), while the mechanism of energy landscape leads to an "S"-shape anomaly with a much larger time scale (~ 10 μs) determined by the strength of the internal field. Analysis of the "S"-shape anomaly reveals the key parameters of the energy landscape and the switching dynamics.

Epitaxial heterostructures h-$YbFeO_3$/$CoFe_2O_4$ (CFO)/$La_{0.7}Sr_{0.33}MnO_3$ (LSMO) were grown on $SrTiO_3$ (STO) (111) substrate by pulsed laser deposition. The CFO layer is to buffer the lattice mismatch between h-$YbFeO_3$ and LSMO bottom electrode[23,24]; its impact as dielectric layer on improper ferroelectricity can be minimized when h-$YbFeO_3$ is thick enough[23]. In this study, we keep the thickness of the CFO and LSMO layers at 10 nm and 30 nm, respectively. X-ray diffraction (XRD) $\theta - 2\theta$ scan, reflection high energy electron diffraction (RHEED), and high-angle annular dark field (HADDF) STEM confirm the $P6_3cm$ structure of h-$YbFeO_3$. (Section S3 within the Supplemental Material) The piezoresponse force microscopy (PFM) measurements on bare surface and the polarization-voltage ($P$-$V$) hysteresis loops measured using the PUND (positive up and negative down) method demonstrates the switchable spontaneous polarization (Section S3 within the Supplemental Material) .

The polarization switching of h-$YbFeO_3$ is measured using a RC circuit with a square wave of source voltage ($V_S$) (**Fig. 2(a)**). Here the resistor limits the currents and polarization switching speed, potentially allowing the ferroelectrics to gradually traverse the energy landscape. **Fig. 2(b)** shows the waveform of $V_S$ and $V_{FE}$ which is the voltage on the ferroelectric capacitor. In contrast to the normal charging process of capacitor in which $dV/dt$ constantly decreases and remains positive, two anomalous features of $dV_{FE}/dt$ can be identified here (**Fig. 2(b)**). The first anomaly occurs in the early stage of ~1 μs time scale. As shown in **Fig. 2(b)** inset, $V_{FE}$ waveform is non-monotonic with a segment of $dV_{FE}/dt < 0$. Since the current $I = (V_S - V_{FE})/R$ is always positive for $V_S > 0$, one has $dq/dt = I > 0$, where $R$ is the resistance of the resistor in **Fig. 2(a)**. Hence, $dV_{FE}/dt < 0$ corresponds to a dynamic NC due to $dq/dV_{FE} = (dq/dt)/(dV_{FE}/dt) < 0$. The second anomaly appears at a later stage of ~ 10 μs time scale. Here $dV_{FE}/dt$ remains positive but exhibits a minimum and a subsequent maximum, as shown in **Fig. 2(c)**, as expected by the LK model[15] (Section S1.3 within the Supplemental Material). We call this feature "S"-shape anomaly. (Both anomalies are also observed in h-$ScFeO_3$/LSMO, see Section S4&S5 within the Supplemental Material)

Next, we discuss the relationships between the first anomaly (NC) and the nucleation mechanism. As shown in **Fig. 3(a)**, we have studied the switching dynamics using square-pulse waveform with the h-$YbFeO_3$ capacitor directly connected to the voltage source[23]. In the early stage, the ferroelectric needs to overcome the nucleation energy barrier, which is responsible for a small initial switching speed $dP/dt$. Near the end of the nucleation stage, $|dP/dt|$ reaches a peak at around ~1 μs before $P$ saturates at a long time. According to Kirchhoff's rule and the approximation $A\, dP/dt \approx - dq/dt$ (Section S6 within the Supplemental Material), one has

$$V_{FE} = V_S - IR = V_S + A \frac{dP}{dt} R, \tag{1}$$

where $A$ is the area of the capacitor. Eq. (1) suggests that the $|dP/dt|$ peak causes a dip in $V_{FE}$ (note for switching $V_S$ from negative to positive, $dP/dt < 0$) and the NC, as observed in **Fig. 2(b)** inset. The switching process of the h-$YbFeO_3$ capacitor can be described within the framework of nucleation-limited switching (NLS) model[28] with a distribution of nucleation time $t_0$[29-31]. The voltage dependence of characteristic switching time $\log_{10}(t_0)$ and the width of distribution functions $w$ are displayed in **Fig. 3(b)**. The ~1 μs time scale of matches that of NC observed in **Fig. 2(b)**.

In principle, nucleation time decreases with external voltage, which is consistent with the observation in **Fig. 3(c)**, which shows the $V_{FE}$ early-stage waveform measured with $R$ = 6 kΩ at different $V_S$ as a two-dimensional image. A $V_{FE}$ dip accompanied by NC is present for all $V_S$, as indicated by the calculated $t_{dip}$ (black dots). Overall, $t_{dip}$ decreases as $V_S$ increases as expected for nucleation time. The weak dependence of $t_{dip}$ on $V_S$ may have to do with the incomplete switching for smaller $V_S$ that reduces the nucleation time. A circuit model predicts that the current peak (corresponding to $V_{FE}$ dip) caused by nucleation is shifted earlier unless the RC time constant is much larger than $t_0$ (Fig. S2.2 within the Supplemental Material). This agrees with **Fig. 3(d)**, where $t_{dip}$ increases with $t_{RC} \equiv RC$ and saturates at large $t_{RC}$ values, $C$ is the capacitance of the ferroelectric capacitor. The model also predicts that $t_{dip}$ saturates at 0.61 $t_0$ for $t_{RC} >> t_0$. One can calculate $t_0 = 1.4 \pm 0.1$ μs from **Fig. 3(d)**, which agrees with the results in **Fig. 3(a&b).**

We then examine the correlation between the "S"-shape anomaly and the free-energy landscape mechanism described by the LK model. According to the LK model, the polarization switching is driven by the sum of the internal field $E_{LK} = -\frac{\partial g}{\partial P}|_{E_{FE}=0}$ and the external field $E_{FE} = -\frac{V_{FE}}{t_{FE}}$ as:

$$\frac{dP}{dt} = \rho(E_{LK} + E_{FE}) \tag{2}$$

where $g$ is the Gibbs free energy density and $t_{FE}$ is the thickness of the ferroelectric, $\rho$ is the viscosity coefficient. Combining Eq. (1) and Eq. (2), one finds the direct effect of $E_{LK}$ on d$P$/d$t$ and $V_{FE}$:

$$V_{FE} = V_S + \frac{1}{\frac{t_\rho}{t_{RC0}}+1}(V_S - t_{FE}E_{LK}) \tag{3}$$

$$\frac{dP}{dt} = \frac{\varepsilon_0}{t_\rho + t_{RC0}}(E_{LK} + E_S) \tag{4}$$

where $t_\rho \equiv \frac{\varepsilon_0}{\rho}$ is the timescale related to viscosity, $t_{RC0} \equiv RC_0$, $C_0 \equiv \frac{A\varepsilon_0}{t_{FE}}$, $E_S \equiv -\frac{V_S}{t_F}$. Eq. (3) suggest non-monotonic $E_{LK}$ causes non-monotonic waveform of $V_{FE}$.

Two implications about polarization switching timescale from Eq. (4) can be verified by the behavior of the observed "S"-shape anomaly. Eq. (4) suggests that $E_S$ needs to be large enough to exceed the peak value of $E_{LK}$ (**Fig. 1(a)**) and thus overcome the energy barrier of polarization switching from one minimum to the other. **Fig. 4(a)** shows the d$V_{FE}$/d$t$ waveform measured with $R$ = 6 kΩ at various $V_S$ as a two-dimensional image. The "S"-shape anomaly that features a minimum and a subsequent maximum in d$V_{FE}$/d$t$, disappears at a cut-off voltage $V_{Scut} = 2.5\pm0.5$ V, in contrast to the presence of NC for all voltages in **Fig. 3(c)**. Meanwhile, $t_{max}$, the time when the maximum of d$V_{FE}$/d$t$ occurs, which is plotted on top of the image (black dots), appears to diverge at $V_{Scut}$. In addition, $V_{Scut}$ appears to be independent of $R$ (**Fig. 4(b)**). Indeed, according to Eq. (4), polarization switching timescale diverges when $E_S$ approaches the peak value of $E_{LK}$ and this is independent of $R$. Hence, the peak value of $E_{LK}$ is found to be approximately $V_{Scut}/t_{FE} \approx (4\pm1)\times10^7$ V/m for h-$YbFeO_3$ thin films, comparable to the observed coercivity[23]. Eq. (4) also suggests that polarization switching timescale is proportional to $t_{RC0}+t_\rho$, which is indeed observed in **Fig. 4(c)**. Hence, $t_\rho$ can be found from the horizontal intercept of the $t_{max}(t_{RC0})$ relation in **Fig. 4(c)**; the results are independent of $V_S$ considering the experimental uncertainty, as shown in **Fig. 4(d)**. Overall, we found $t_\rho$ = 21±5 ns, corresponding to ρ =(4±1)×$10^{-4}$ F/(m·s) for h-$YbFeO_3$, which is of similar order of magnitude to that reported for ferroelectric $HfO_2$.[15] Analysis assuming proper ferroelectrics indicate that the slope of the $t_{max}(t_{RC0})$ relation, or the magnitude of $t_{max}$ depends on the peak value of $E_{LK}$ (Section S1.3.1 within the Supplemental Material).

Unlike the homogenous switching assumed by the classic LK model, polarization switching is normally inhomogeneous involving local nucleation and domain wall motion. Nevertheless, local polarization switching is still expected to follow the free-energy landscape decorated with the domain wall energy[32]. Hence, a convolution of the effects of the free-energy landscape and inhomogeneity is expected, which smears the modulation of $V_{FE}$ waveform by $E_{LK}$ described by Eq. (3). As a result, NC response may turn into the "S"-shape anomaly or even totally vanish. The shape of $E_{LK}$ ($P$) is key for the observation of "S"-shape anomaly. Thus, we further discuss the abrupt changes of $E_{LK}$ of the improper ferroelectric hexagonal ferrites, originating from the multi-variable free-energy landscape (Section S6.3 within the Supplemental Material).

Here we adopt a simplified Gibbs free energy density for hexagonal ferrites assuming a constant $Q$, since the structural distortion has a larger energy scale than that of the electrostatic interactions:

$$g(\phi, \mathrm{P}) = g_0 - (a_P \cos 3\phi + E_F)P + \frac{b_\mathrm{P}}{2}P^2 - \frac{\varepsilon_0}{2}E_F^2 \quad (5)$$

where $g_0$, $a_P$>0 and $b_P$>0 are coefficients that only depend on $Q$, $\varepsilon_0$ is the vacuum permittivity. $g(\phi, \mathrm{P})$ is displayed in **Fig. 4(e)** near two neighboring potential wells at $\phi$ =0 and π/3 respectively, where $P_0 = \frac{a_P}{b_P}$ is the spontaneous polarization at zero field.

Assuming an initial state of energy minimum with $\phi$ = 0 and $P = P_0$ >0. In step 1 (**Figs. 5(a&b)**), a constant $E_{FE}$ <0 is applied to reduce $P$ from $P_0$. Because $P < P_0$, $E_{LK} = a_P - b_P P > 0$ resists $E_{FE}$. In addition, $\phi$ = 0 remains a local minimum because $\partial^2 g/\partial\phi^2 = 9a_P P\cos(3\phi)>0$. In step 2, $P$ becomes negative. $\phi$ = 0 becomes a maximum because $\partial^2 g/\partial\phi^2$ <0, which triggers a sudden change of $\phi$. Minimizing $g$ with respect to $\phi$ results in $g = g_0 - 1/2\, b_P P^2 - 1/2\, \varepsilon_0 E_{FE}^2$, corresponding to $E_{LK} = b_P P < 0$. Here $E_{LK}$ is in the same direction as $E_{FE}$ which accelerates polarization switching. In step 3, the system reaches $\phi$ = π/3 (global minimum for $E_{FE}$ <0). Polarization switching adopts a constant $\phi$. Here $E_{LK} = a_P - b_P P$ becomes positive again, which resists the polarization switching. Although the overall trend, i.e., one minimum and one maximum of $E_{LK}$ in **Fig. 4(f)**, is similar to that of proper ferroelectric in **Fig. 1(a)**, the more abrupt change of $E_{LK}(P)$ in **Fig. 4(f)** may make its effect potentially more distinguishable.

We use a hybrid model to account for the convoluted effects of free-energy landscape and nucleation (inhomogeneity) quantitatively. Starting with the NLS model[23,29-31] in which $|dP/dt|$ is determined by $E_{FE}$ and the nucleation processes, we replace $E_{FE}$ with $E_{FE} + E_{LK}$ so that $|dP/dt|$ is determined by $E_{FE} + E_{LK}$ instead, which is the foundation of the LK model described by Eq. (2). Since the essential shapes of $E_{LK}$ in **Fig. 4(f)** and **Fig. 1(a)** are similar, to facilitate the numerical simulation, we borrow the polynomial form of $E_{LK}$ from the proper FE and assume

$$E_{\mathrm{LK,eff}}(P) = -2\alpha P - 4\beta P^3 \quad (6)$$

where $\alpha$ and $\beta$ are fitting parameters, similar to the Landau coefficients.

As shown in **Fig. 5(a)**, the theoretical fitting replicates the $V_{FE}$ waveform, achieving much higher consistency in whole time range compared with previous work[12,14,15,33,34]. For the NC region (**Fig. 5(b)**), once $V_S$ changes sign, the nucleation process causes a maximum of $|dP/dt|$ (see **Fig. S6.2**), a minimum in $V_{FE}$, and the corresponding NC. Noticeably, $V_{LK,eff}$ at this time scale shows a monotonic decrease, which is clearly unrelated to NC, whereas, for the "S" shape region, $V_{LK,eff}$ exhibits non-monotonic shape (**Fig. 5(c)**). By contrast, $|dP/dt|$ (see **Fig. S10**) has an "S" shape, i.e., monotonic but with a minimum and a maximum in the slope, which explains the "S"-shape anomaly in the $V_{FE}$ waveform according to Eq. (1). The shape of $|dP/dt|$ is damped since it is determined by $V_{LK,eff} + V_{FE}$ instead of $V_{FE}$ in the hybrid model.

In summary, the dual mechanisms for transient NC anomaly, including nucleation and free-energy landscape, are identified as the origin of the early-stage transient NC and the later-stage "S"-shape anomaly, respectively, and well fit by a hybrid model, in improper ferroelectric hexagonal ferrites. The discernable manifestation of the free-energy landscape can be attributed to the abrupt change of LK field, inherited from the multi-variable energy landscape. The dependence of the time scale of NC and that of the "S" shape on RC time constant and external field allows extraction of the intrinsic parameters of nucleation, free-energy landscape, and viscosity. These results may settle the long-standing debate on the origin of transient NC and suggest the potential of hexagonal ferrites for practical applications in low-power electronic devices.

### Acknowledgment:

This work is primarily supported by the National Science Foundation (NSF), Division of Materials Research (DMR) under Grant No. DMR-1454618. The research is performed in part in the Nebraska Nanoscale Facility: National Nanotechnology Coordinated Infrastructure and the Nebraska Center for Materials and Nanoscience, which are supported by the NSF under Grant No. ECCS- 1542182, and the Nebraska Research Initiative. Work at Washington University is supported by NSF Grant No. DMR-1806147. STEM experiments are conducted at the Center for Nanophase Materials Sciences at Oak Ridge National Laboratory, which is a Department of Energy (DOE) Office of Science User Facility, through a user project. This research used resources of the Advanced Photon Source, a U.S. Department of Energy (DOE) Office of Science user facility operated for the DOE Office of Science by Argonne National Laboratory under Contract No. DE-AC02-06CH11357.

## References

1. Landau, L. D. & Lifshitz, E. M. Electrodynamics of Continuous Media. (Oxford, 1960).

2. Ong, L.-H., Osman, J. & Tilley, D. R. Landau theory of second-order phase transitions in ferroelectric films. Phys. Rev. B 63,144109 (2001).

3. Daraktchiev, M., Catalan, G. & Scott, J. F. Landau Theory of Ferroelectric Domain Walls in Magnetoelectrics. Ferroelectrics 375, 122-131 (2008).

4. Kumar, A. & Waghmare, U. V. First-principles free energies and Ginzburg-Landau theory of domains and ferroelectric phase transitions in $BaTiO_3$. Phys. Rev. B 82,054117(2010).

5. Íñiguez, J., Zubko, P., Luk'yanchuk, I. *et al.* Ferroelectric negative capacitance. *Nat Rev Mater* **4**, 243–256 (2019)

6. Gruverman. A., Rodriguez. B. J., Dehoff. C. Direct studies of domain switching dynamics in thin film ferroelectric capacitors. Appl. Phys. Lett. 87, 082902 (2005).

7. D. J. Jung , M. Dawber , J. F. Scott. et al. Switching Dynamics in Ferroelectric Thin Films: An Experimental Survey. Integrated Ferroelectrics, 48:1, 59-68 (2002).

8. J. F. Scott. A review of ferroelectric switching. Ferroelectrics, 503:1, 117-132(2016).

9. Gruverman. A, Kholkin. A . Nanoscale ferroelectrics: processing, characterization and future trends. Rep. Prog. Phys. 69, 2443(2006).

10. Vizdrik. G., Ducharme. S., Fridkin. V. M. et al. Kinetics of ferroelectric switching in ultrathin films. Phys. Rev. B 68, 094113(2003).

11. Fridkin. V., Levlev. A., Verkhovskaya. K. Switching in One Monolayer of the Ferroelectric Polymer, Ferroelectrics, 314:1, 37-40(2005).

12. Khan. A., Chatterjee. K., Wang. B. et al. Negative capacitance in a ferroelectric capacitor.Nature Mater 14, 182–186 (2015).

13. Hoffmann, M., Fengler, F.P.G., Herzig, M. et al. Unveiling the double-well energy landscape in a ferroelectric layer. Nature 565, 464–467 (2019).

14. Hoffmann, M., Pešić, M., Chatterjee, K., Khan, A. I., Salahuddin, S., Slesazeck, S., Schroeder, Uwe., Mikolajick, T. Direct Observation of Negative Capacitance in Polycrystalline Ferroelectric $HfO_2$. Adv. Funct. Mater. 26, 8643–8649 (2016).

15. Chang. S-C., Avci.U. E., Nikonov. D. E. Physical Origin of Transient Negative Capacitance in a Ferroelectric Capacitor. Phys. Rev. Applied 9, 014010 (2018).

16. Yang. T. J., Gopalan. V., Swart. P. J. et al. Direct Observation of Pinning and Bowing of a Single Ferroelectric Domain Wall. Phys. Rev. Lett. 82, 4106 (1999).

17. Shin, YH., Grinberg, I., Chen, IW. et al. Nucleation and growth mechanism of ferroelectric domain-wall motion. Nature 449, 881–884 (2007).

18. Grigoriev. A., Do. D.H, Kim. D. M. et al. Nanosecond Domain Wall Dynamics in Ferroelectric $Pb(Zr,Ti)O_3$ Thin Films. Phys. Rev. Lett. 96, 187601 (2006).

19. Catalan, G., Jiménez, D., Gruverman, A. Negative capacitance detected. Nature Mater 14, 137–139 (2015).

20. Wang, W. et al. Room-temperature multiferroic hexagonal $LuFeO_3$ films. Phys. Rev. Lett. 110, 237601 (2013).

21. Xu, X. & Wang, W. Multiferroic hexagonal ferrites (h-$RFeO_3$, R = Y, Dy-Lu): a brief experimental review. Mod. Phys. Lett. B 28, 1430008 (2014).

22. Sinha, K. et al. Tuning the Neel Temperature of Hexagonal Ferrites by Structural Distortion. Phys. Rev. Lett. 121, 237203 (2018).

23. Yun, Y. et al. Spontaneous Polarization in an Ultrathin Improper-Ferroelectric/Dielectric Bilayer in a Capacitor Structure at Cryogenic Temperatures. Phys. Rev. Appl. 18 (2022).

24. Li, X., Yun, Y., Thind, A.S. et al. Domain-wall magnetoelectric coupling in multiferroic hexagonal $YbFeO_3$ films. Sci Rep 13, 1755 (2023).

25. Li, X., Yun, Y., Xu, X. S. Improper ferroelectricity in ultrathin hexagonal ferrites films. Appl. Phys. Lett. 122, 182901 (2023)

26. Artyukhin, S., Delaney, K. T., Spaldin, N. A. & Mostovoy, M. Landau theory of topological defects in multiferroic hexagonal manganites. Nature. Mater. 13, 42–49 (2014).

27. Zhang. C. X., Yang. K. L., Jia. P. et al. Effects of temperature and electric field on order parameters in ferroelectric hexagonal manganites. J. Appl. Phys. 123, 094102 (2018).

28. Hoffmann, M., Slesazeck, S., Schroeder, U. et al. What’s next for negative capacitance electronics?. Nat Electron 3, 504–506 (2020).

29. Tagantsev, A. K., Stolichnov, I., Setter, N., Cross, J. S. & Tsukada, M. Non-Kolmogorov-Avrami switching kinetics in ferroelectric thin films. Phys. Rev. B 66, 214109 (2002).

30. Jo. J. Y., Han. H. S., Yoon. J.-G. Domain Switching Kinetics in Disordered Ferroelectric Thin Films. Phys. Rev. Lett. 99, 267602 (2007).

31. Gruverman. A., Wu. D., and J. F. Scott. Piezoresponse Force Microscopy Studies of Switching Behavior of Ferroelectric Capacitors on a 100-ns Time Scale. Phys. Rev. Lett. 100, 097601(2008).

32. Yoshihiro Ishibashi. A Model of Polarization Reversal in Ferroelectrics. J. Phys. Soc. Japan 59 4148 (1990).

33. Kim. Y. J., Park. H. W., Hyun. S. D. et al. Voltage Drop in a Ferroelectric Single Layer Capacitor by Retarded Domain Nucleation. Nano Lett 17, 7796-7802(2017).

34. Hao. Y., Li. T., Yun. Y. et al. Tuning Negative Capacitance in $PbZr_{0.2}Ti_{0.8}O_3/SrTiO_3$ Heterostructures via Layer Thickness Ratio. Phys. Rev. Applied 16, 034004(2021).

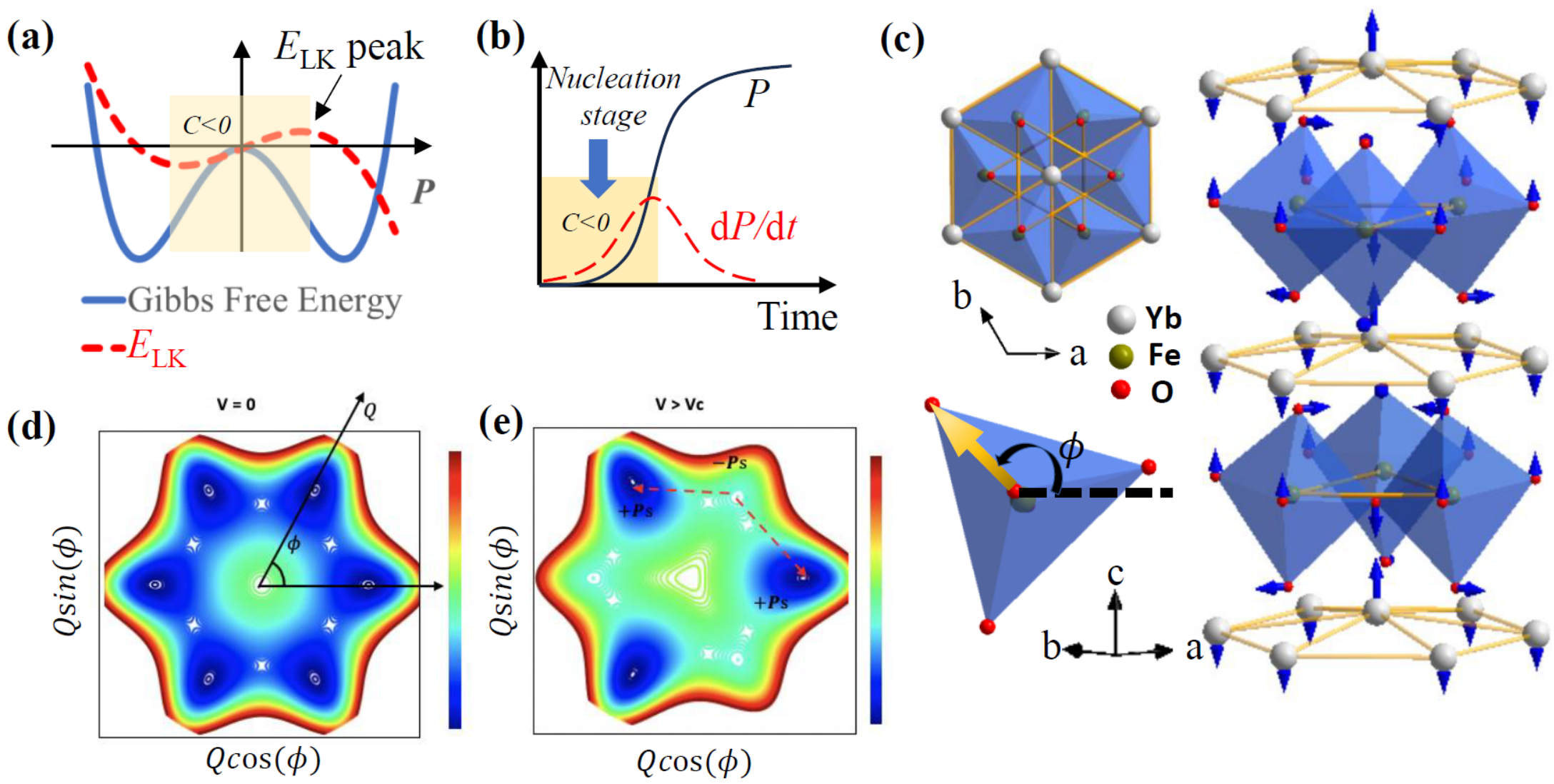

**Fig. 1** (a) Schematics of free-energy landscape and internal field $E_{LK}$ as well as the NC region for proper ferroelectrics. (b). Time-dependent polarization and the NC region for a nucleation process. (c) Atomic structure of h-$YbFeO_3$. The arrows indicate the displacement pattern of the $K_3$ distortion mode. (d) The free-energy landscape of improper ferroelectric h-$RFeO_3$ with (d) zero and (e) positive external field.

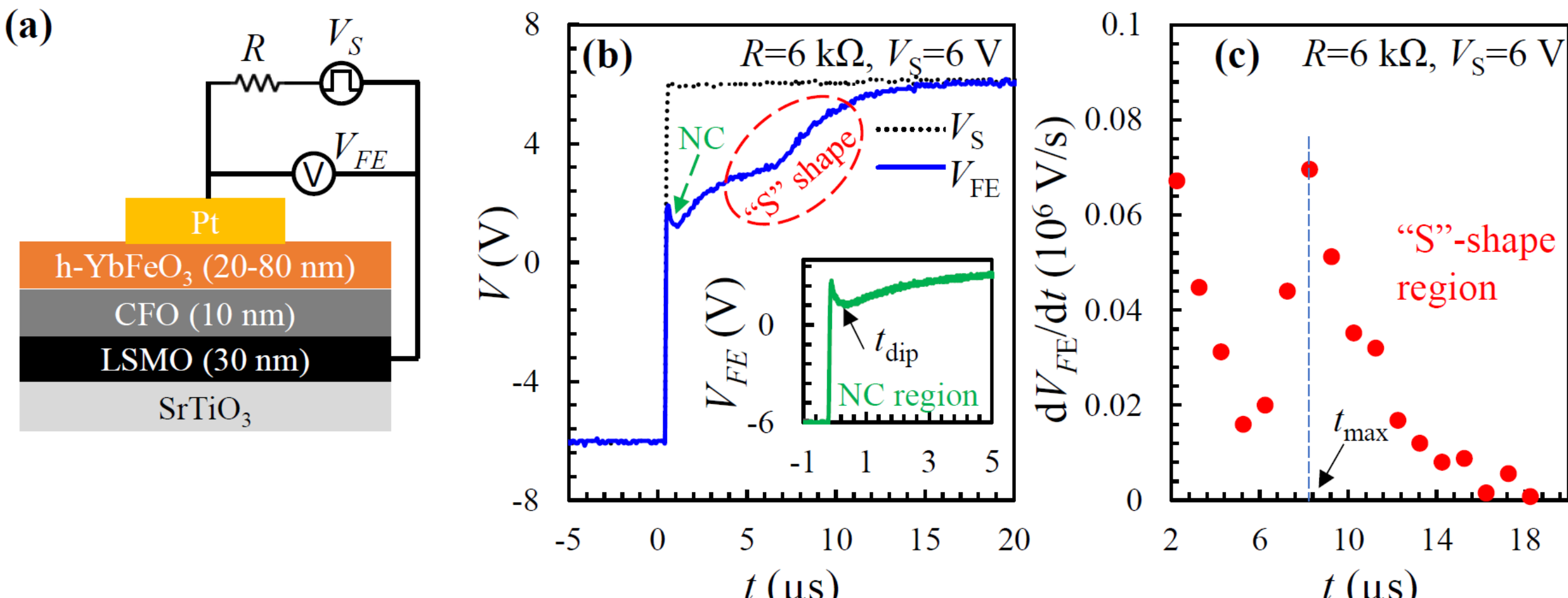

**Fig. 2** (a) Schematic diagram of experimental setup. $V_{FE}$ (b) and d$V_{FE}$/d$t$ (c) waveform measured with $R$=6 kΩ and $V_S$ = 6V. The sparse data points of d$V_{FE}$/d$t$ in (c) is the result of large bin size of the $V_{FE}$ data chosen to calculate the derivative. The thickness of h-YbFeO$_3$ is 60 nm.

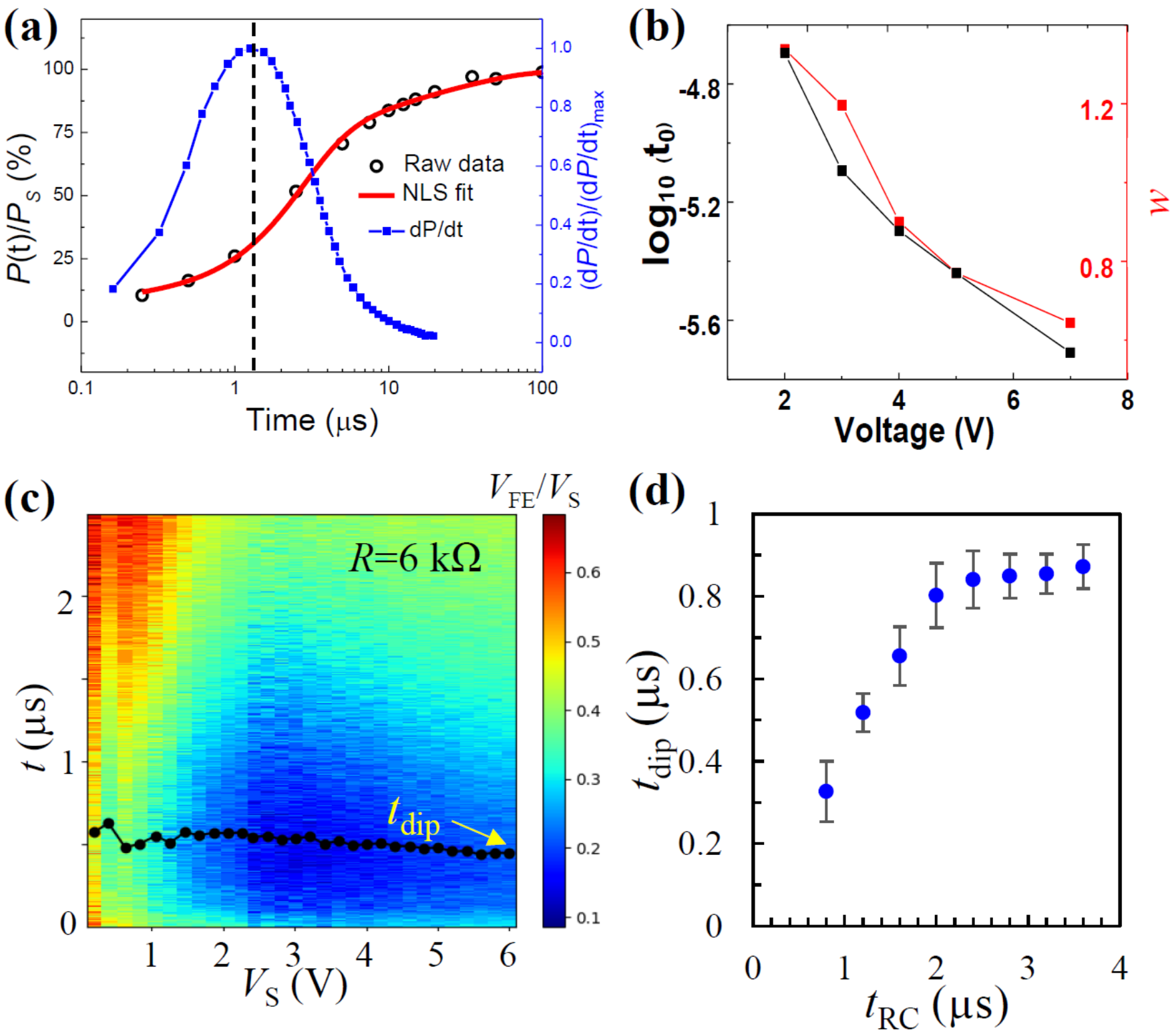

**Fig. 3** (a) Time-dependent polarization switching and related fitting by the NLS model, with data from Ref. 23. (b) Dependence of nucleation time $t_0$ and distribution width $w$ on the applied voltage. (c) $V_{FE}$ waveform measured with $R$=6 kΩ at different $V_S$. The dots indicate the time $t_{dip}$ where the minimum occurs. (d) Dependence of $t_{dip}$ on $t_{RC}$ where $C \approx 0.2$ nF. For each $t_{RC}$ (or $R$) value, the $t_{dip}$ value and error bar are calculated from the average and standard deviation of the values measured with different $V_S$ respectively. The thickness of h-$YbFeO_3$ is 31 nm for (a) and (b) and 60 nm for (c) and (d).

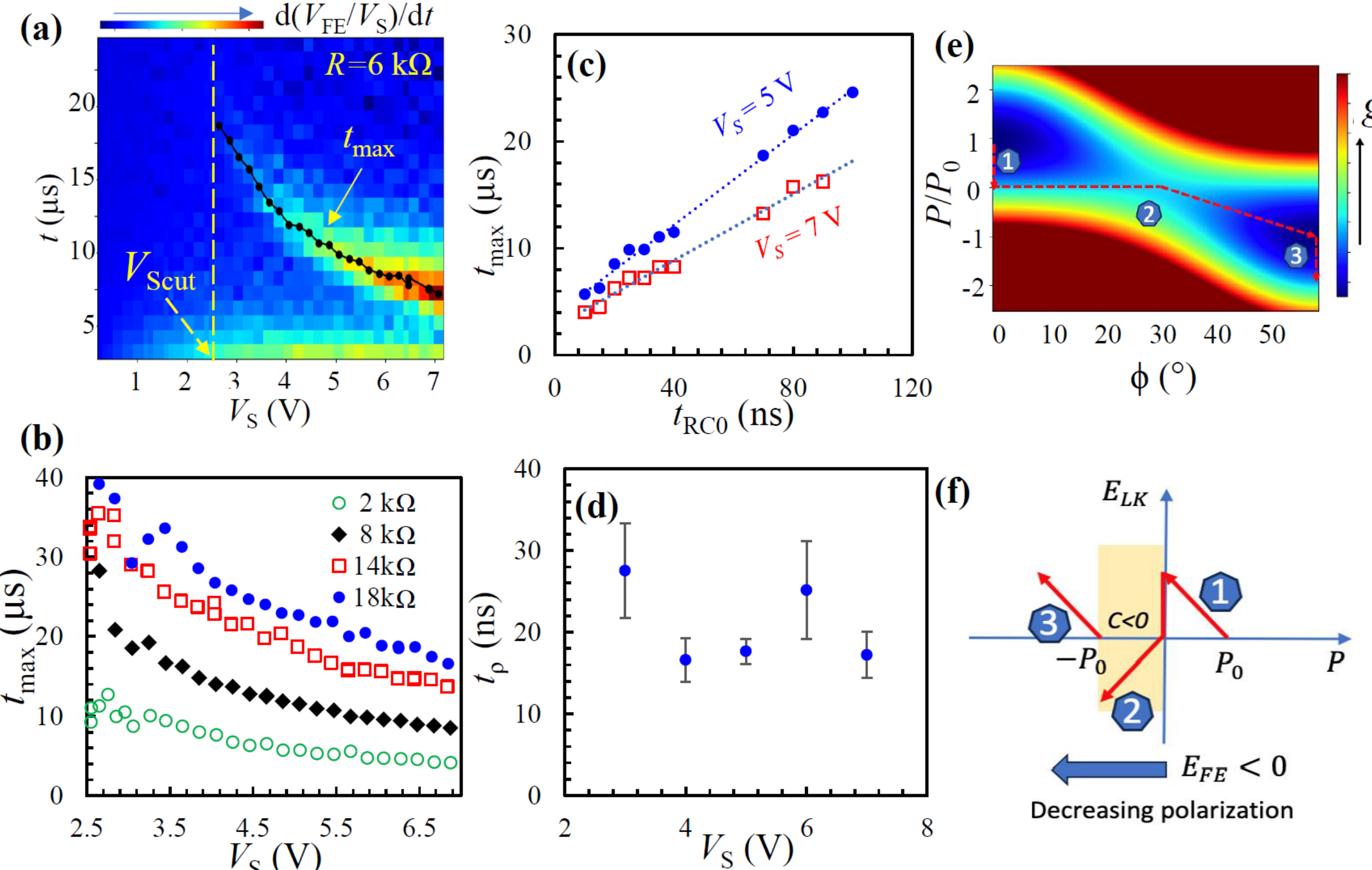

**Fig. 4** (a) d$V_{FE}$/d$t$ waveform measured with $R$=6 kΩ at different $V_S$. The dots indicate the time $t_{max}$ where the maximum occurs. Vertical dashed line indicates a cut-off voltage $V_{Scut}$ at which the maximum disappears. (b) Dependence of $t_{max}$ on $V_S$ for different $R$. (c) Dependence of $t_{max}$ on $t_{RC0}$ where $C_0 \approx 5$ pF. Markers (round and square) and dashed lines are experimental data and linear fit respectively. (d) $t_\rho$ extracted from (c), see text. Schematic of the polarization switching path (e) and the polarization-dependent $E_{LK}$ (f) from -$P_0$ to +$P_0$ for improper ferroelectric h-$R$FeO$_3$. The thickness of h-YbFeO$_3$ is 60 nm.

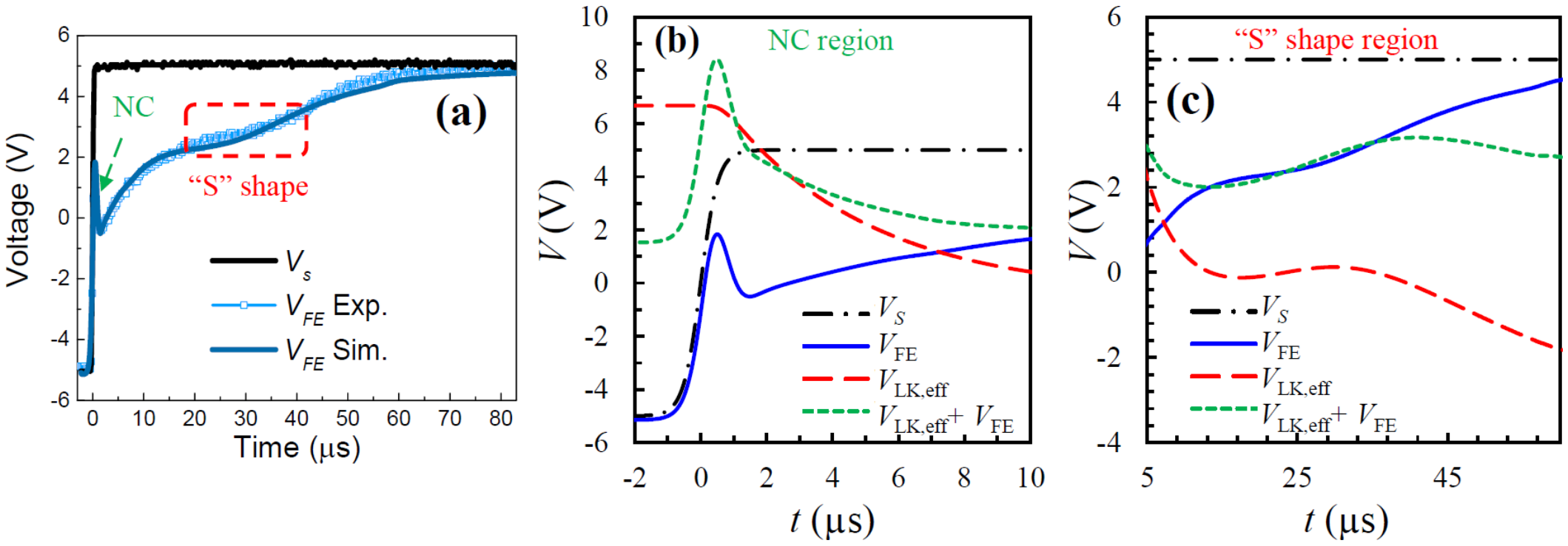

**Fig. 5** (a) Experimental and simulated waveform of $V_{FE}$. Simulated waveform of $V_S$, $V_{FE}$, $V_{LK,eff}$, and $V_{FE}$+ $V_{LK,eff}$ in the NC region (b) and the "S" shape region (c). The thickness of h-$YbFeO_3$ is 60 nm.